\documentclass[twocolumn,showpacs,preprintnumbers,amsmath,amssymb]{revtex4}
\def\beneq{\begin{equation}}
\def\eneq{\end{equation}}
\def\bea{\begin{eqnarray}}
\def\eea{\end{eqnarray}}
\usepackage{graphicx}
\usepackage{dcolumn}
\usepackage{bm}

\begin{document}
\title{Interplay of instabilities in mounded surface growth}
\author{Buddhapriya Chakrabarti}
\email{buddho@physics.umass.edu} \altaffiliation[Now
at]{Department of Physics, University of Massachusetts, Amherst,
MA 01003.}
\author{Chandan Dasgupta}
\email{cdgupta@physics.iisc.ernet.in}
\affiliation{Centre for Condensed Matter Theory, Department of Physics, 
Indian Institute of Science, Bangalore 560012, INDIA.}
\date{\today}
\begin{abstract}
We numerically study a one-dimensional conserved growth equation
with competing linear (Ehrlich-Schwoebel) and nonlinear
instabilities. As a control parameter is varied, this model exhibits a
non-equilibrium phase transition between two mounded states, one of which 
exhibits slope selection and the other does not. 
The coarsening behavior of mounds in these two phases is studied
in detail. In the absence of noise
the steady-state configuration depends crucially on which of the
two instabilities dominates the early time behavior. 
\end{abstract}
\maketitle

The phenomenon of formation and 
coarsening of mounds in epitaxially grown thin films~\cite{Krug:97} is a subject
of much recent experimental~\cite{Johnson:94,Apostolopoulos:00} and 
theoretical~\cite{Siegert:94,Politi:96,Rost:97} interest.
Traditionally, the formation of mounds has been
attributed to the presence of an Ehrlich-Schwoebel (ES)
step-edge barrier \cite{Ehrlich:96,Schwoebel:69} that hinders the downward motion of
atoms across the edge of a step. 
This leads to an effective
``uphill'' surface current \cite{Villain:91} that has a destabilizing
effect, leading to
the formation of mounded structures. The ES mechanism is usually represented
in continuum growth equations as a {\it linear} instability~\cite{Villain:91}
that is controlled by higher-order nonlinear terms. In ES-type models, the
characteristics of the coarsening process, in which the typical lateral 
size of the mounds grows 
in time, depends on the details of the model. In these models, slope selection
occurs (i.e. the slope of the mounds remains constant during coarsening)
only if the ``ES part'' of the slope dependent surface current has one or more
stable zeros as a function of the slope.  

In our
earlier work~\cite{Chakrabarti:03,Chakrabarti:04} on a class of spatially discrete, 
conserved, one-dimensional (1d) models of epitaxial growth, we found 
a mechanism of mound formation and coarsening with slope selection  
that is different from the conventional
ES mechanism. We studied the spatially discretized Lai-Das Sarma equation
\cite{Lai:91} of MBE growth and an atomistic model~\cite{Kim:94} that provides a
discrete realization of the dynamics described by this equation, 
and found the occurrence of a
{\it nonlinear} instability~\cite{Dasgupta:97} in which isolated pillars 
or grooves grow
in time if their height or depth exceeds a ``critical'' value. When this
instability is
controlled by the introduction of 
an infinite number of higher order gradient
nonlinearities, these models show, for a range of parameter values, 
the formation of mounds with a 
well-defined slope that
remains constant during the coarsening process. These results demonstrate that
mound formation and power-law coarsening with slope selection can occur in
the absence of an ES instability.

In most experimentally studied systems, however, it is believed
that the ES step-edge barrier is present, although it may possibly be
very weak. It is, therefore, important to understand how the behavior of
our models would be modified when the ES mechanism is incorporated in their
kinetics. To address this issue, we have studied, using numerical integration, 
a spatially discretized 1d growth equation in which a linear ES-type
instability is present in conjunction with the nonlinear
instability mentioned above. The main results of this study are as follows.
Depending on the values of the parameters appearing in the model, we find two
different mounded states. If the parameters are such that the nonlinear 
instability is the dominant one, then the behavior of the system 
is similar to that found in our earlier studies~\cite{Chakrabarti:03,Chakrabarti:04}: it
exhibits the formation of triangular mounds 
and power-law coarsening with slope selection. If, on
the other hand, the linear instability is dominant, then the
system exhibits a different kind of mounded state in which the mounds
have a cusp-like shape and they steepen during the coarsening process.
We call these two mounded states ``faceted'' and ``cusped'', respectively. 
As the parameters are changed, the system undergoes a {\it dynamical phase
transition} from one of these mounded states to the other.

We study a spatially discretized 1d version of 
the fourth order conserved growth equation, proposed in
the context of MBE growth \cite{Lai:91,Villain:91}, in which
the nonlinear instability \cite{Dasgupta:97}
is controlled using a control function \cite{Chakrabarti:03,Chakrabarti:04} of the form 
proposed by
Politi and Villain
\cite{Politi:96}
and a linear instability of the form proposed by Johnson 
{\it et al.} \cite{Johnson:94} to
represent the ES effect is also included. Thus, the equation of
motion of the interface height in appropriately
non-dimensionalized form is written as:
\begin{eqnarray}
\partial h_i/\partial t = -\tilde{\nabla}^4 h_i +
\lambda \tilde{\nabla}^2 (| \tilde{\nabla} h_i |^2/(1 + c_1 |\tilde{\nabla} h_i |^2)) - \nonumber \\ \tilde{\nabla}(\tilde{\nabla} h_i/(1+c_2 |\tilde{\nabla} h_i |^2)) + \eta_i(t), \label{pt1_chap4}
\end{eqnarray}
where $h_i$ is the non-dimensionalized height variable at lattice site $i$,
and $\tilde{\nabla}$ and $\tilde{\nabla^2}$ are the lattice versions
of the derivative and Laplacian operators, respectively, calculated using the
nearest neighbors as outlined in our earlier 
papers\cite{Chakrabarti:03,Chakrabarti:04}. In Eq.(\ref{pt1_chap4}), $c_1$, $c_2$ and $\lambda$ 
are constants (model parameters), and $\eta_i(t)$ represents uncorrelated 
random noise with zero mean and unit variance. 
Our results are based
on numerical integration of this equation in 1d,
using a simple Euler scheme\cite{Dasgupta:97} in which the time evolution of
the height variables is given by
\begin{eqnarray}
h_i(t + \Delta t) = h_i(t) + \Delta t \{\tilde{\nabla}^{2} [ -
\tilde{\nabla}^{2} h_i(t) + \nonumber \\ \lambda (|\tilde{\nabla}
h_i(t)|^2/(1+ c_1|\tilde{\nabla} h_i(t)|^2))] - \nonumber \\
\tilde{\nabla}(\tilde{\nabla} h_i(t)/(1+c_2|\tilde{\nabla}
h_i(t)|^2))\} + \sqrt{\Delta t}\, \eta_i(t). \label{pt_chap4}
\end{eqnarray}
In all our calculations, the value of
the parameter $c_2$ was held fixed at unity, and $c_1$ and
$\lambda$ were varied.
Our results are obtained for different system sizes, $40 \le L \le
1000$, with periodic boundary conditions. We do not find any
significant $L$-dependence. We have used a time step $\Delta
t$=$0.01$ for most of our studies and checked that very similar
results are obtained with smaller values of $\Delta t$. The
results obtained by using a more sophisticated
algorithm~\cite{Shampine:94} closely match with those of our Euler
scheme for a small enough $\Delta t$. We have also checked that
the basic features of our results remain unaltered if other boundary conditions
(such as ``fixed'' and ``zero-flux'' boundary conditions considered in
our earlier work~\cite{Chakrabarti:03,Chakrabarti:04}) are used.

\begin{figure}
\includegraphics[width=8cm,height=5cm]{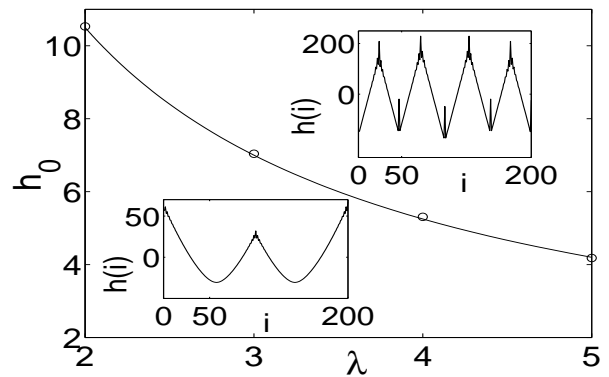}
\caption{\label{fig2chap4} Dependence of the morphology on the initial
conditions in the absence of noise. Main plot: the critical value
of $h_0$ (see text) as a function of 
the nonlinearity parameter $\lambda$ for $c_1=0.01$.
This critical value is approximately given by $h_0
= 21/\lambda$ (solid line). Upper inset: ``faceted'' morphology for $h>h_0$.
Lower inset: ``cusped'' morphology for $h<h_0$.}
\end{figure}

We first describe the results obtained in the absence of the noise term.
In this case, we find that if the parameter $c_1$ is sufficiently small,
then the long
time behavior of the interface
depends on which of the instabilities dominates at early times. In
order to characterize this, we considered initial configurations
with a single pillar of height $h_0$ on an otherwise flat
interface, and studied the long-time behavior as a function of $h_0$.
We find that when
$h_0$ is sufficiently small, so that the nonlinear instability is not
initiated, the linear
instability dominates the time evolution and the resulting morphology
is mounded without slope selection. If on the other hand $h_0$ is large
enough to seed the nonlinear instability, then
mounds with a ``magic'' slope result. This is illustrated
in Fig.~\ref{fig2chap4} which shows 
typical configurations obtained in the two cases: slope selected mounds
(``faceted'') in situations where the height of the central pillar
is greater than a critical value and mounds without slope
selection (``cusp'' like ) otherwise. Fig.~\ref{fig2chap4} also shows
the dependence of the critical value of $h_0$ on $\lambda$, the strength
of the nonlinearity. We find that this critical value is approximately
given by $A/\lambda$, $A \simeq 21$, with 
a weak dependence on the parameter
$c_1$.

\begin{figure}
\includegraphics[width=8cm,height=5cm]{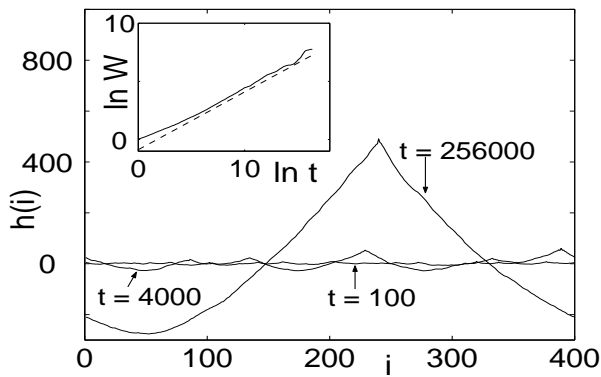}
\caption{\label{fig4chap4} The interface profiles at three
different times ($t$= 100, 4000 and 256000) in a run with noise
starting from a flat state for an $L= 400$  sample with
$\lambda=4.0$ and $c_1=0.05$. A double log plot of the interface
width $W$ as a function of time $t$, 
averaged over $15$ runs for $L= 1000$ samples in
the presence of noise, and a power-law fit are shown in the inset.}
\end{figure}

\begin{figure}
\includegraphics[width=8cm,height=5cm]{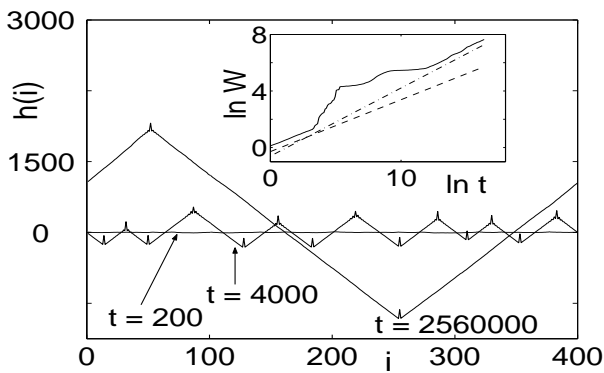}
\caption{\label{fig5chap4} Interface profiles at three different
times ($t$= 200, 4000 and 2560000) in a run starting from a flat
state for an $L= 400$ sample with $\lambda = 4.0$ and $c_1 =
0.01$. A double log plot of the interface width $W$ as a function
of time $t$, 
averaged over $15$ runs for $L= 1000$ samples in the presence of
noise, and power-law fits to the early and late-time data 
are shown in the inset.}
\end{figure}

This ``bistable'' behavior is found only if the control parameter
$c_1$ is
sufficiently small. Otherwise, the system evolves to the
cusped morphology even if the initial configuration has a high
central pillar. Thus,
the faceted morphology is found in the zero-noise
simulations only if $c_1 < c_1^{sp}(\lambda)$ (the superscript $sp$
denotes ``spinodal'', see below) and the initial
configuration is sufficiently rough to seed the nonlinear
instability. This behavior may be understood from a
linear stability analysis. In both faceted and cusped growths,
the finite-sized system evolves, at long times, to a 
profile with a single mound which is a fixed point of the noiseless dynamics.
Examples of such profiles (obtained from simulations with noise which slightly
roughens the profiles) are shown in Figs.~\ref{fig4chap4} and \ref{fig5chap4}.
These fixed points can also be obtained  
by calculating the $h_i$
for which $g_i$=0 for all $i$, where $g_i$ is the term multiplying $\Delta t$
on the right-hand side of Eq.(\ref{pt_chap4}). 
The local stability of the faceted fixed point may be determined from a
calculation of the eigenvalues of the matrix
$M_{ij}=\partial g_i/\partial h_j$ evaluated
at the fixed point. We find that the largest eigenvalue of this matrix crosses
zero at a ``spinodal'' value, $c_1=c^{sp}_1(\lambda)$
(see inset of Fig.~\ref{fig6chap4}), signaling
an instability of the faceted profile.
Thus, for $0 < c_1 < c^{sp}_{1}(\lambda)$,
the dynamics of Eq.(\ref{pt_chap4}) without noise admits two locally stable
invariant profiles: a cusped profile without slope selection, and a faceted one
with slope selection. Depending on the initial
state, the no-noise dynamics takes the system to one of these two fixed
points. For example, an initial state with one pillar on a flat background is
driven by the no-noise dynamics to the cusped fixed point if the height of the
pillar is smaller than a critical value (mentioned earlier), and to the faceted
one otherwise. The dependence of $c^{sp}_{1}$ on the nonlinearity parameter
$\lambda$ is shown in Fig.~\ref{fig6chap4}. Such a ``spinodal line'' does not
exist for the cusped fixed point.

\begin{figure}
\includegraphics[width=8cm,height=5cm]{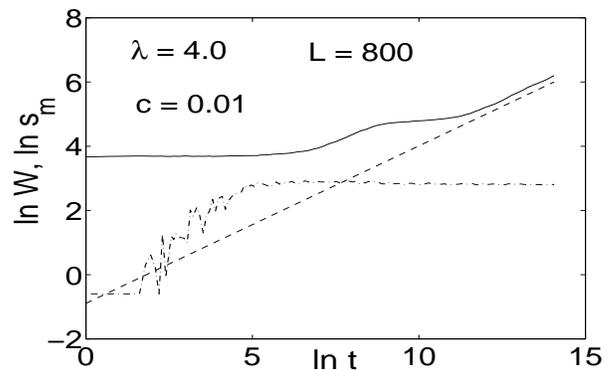}
\caption{\label{fig8chap4} Double log plots of the width $W$ (full
line) and the average of the maximum slope $s_{m}$ (dash-dotted
line) as a function of time $t$, for faceted growth with slope
selection in the absence of noise. This data are for $\lambda =
4.0$, $c_1 = 0.01$, and $L = 800$, averaged
over $20$ initial conditions. In this case the growth exponent
$\beta = 0.5 \pm 0.01$ (the dashed line is the best power-law fit) 
and the steepening exponent $\theta = 0$.}
\end{figure}

\begin{figure}
\includegraphics[width=8cm,height=5cm]{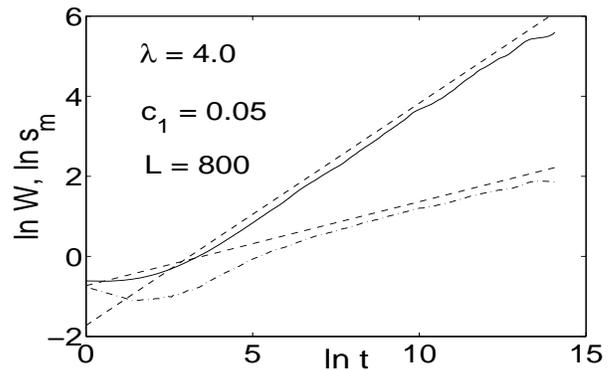}
\caption{\label{fig7chap4} Variation of width $W$ (full line) and 
the average of the maximum
slope, $s_{m}$ (dash-dotted line) as a function of time $t$ on 
a log-log scale for cusped growth
in the absence of noise ($\lambda = 4.0$, $c_1 = 0.05$, 
and $L = 800$ averaged over
$20$ initial conditions). The growth exponent is $\beta
= 0.54 \pm 0.02$, and
the steepening exponent $\theta = 0.2 \pm 0.02$. Best power-law fits are
shown by the dashed lines. }
\end{figure}

\begin{figure}
\includegraphics[width=8cm,height=5cm]{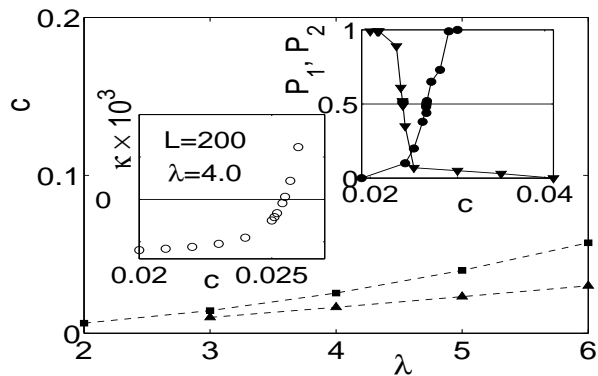}
\caption{\label{fig6chap4} Phase diagram 
for a $L$= 200 system, where the
critical values of the
control parameter $c_1$ are shown as functions of $\lambda$. The critical
value $c_1^{sp}$ ($c_1^{cr}$) 
above which the faceted phase is unstable (metastable) is shown by
squares (triangles). The dashed lines are guides to the eye. Left Inset: Zero
crossing of the largest eigenvalue $\kappa$ of the stability
matrix of the faceted fixed point 
as a function of $c_1$ ($\lambda = 4.0$, $L$ = 200). Right
Inset: The probabilities $P_1$ (circles) and $P_2$ (inverted
triangles) (see text) as functions of $c_1$ with $\lambda = 5.0$
and $L$= 200.}
\end{figure}

We have studied in detail the process of coarsening of the mounds in the
two different (faceted and cusped) growth modes. 
In these simulations, the initial configuration is
obtained by setting $h_i=h_0 r_i$ where $r_i$ is a random number
uniformly distributed between $-1$ and $1$. If 
$c_1 < c_1^{sp}$ and $h_0$ is sufficiently
large to initiate the nonlinear
instability, the system evolves to a faceted structure; a cusped
structure is obtained otherwise. The results reported here were obtained
from averages over 20 such runs with different initial configurations. 
In the region of parameter space, $0 < c < c^{sp}_1 (\lambda)$, where the
faceted phase is locally stable, the mounds coarsen in
time with the slope of the mounds remaining constant. In this case, the
average mound size $R(t)$ is obviously proportional to the interface
width $W(t)$. As shown in Fig~\ref{fig8chap4}, the interface width in 
this growth mode increases as a power-law with time in the long-time
coarsening regime, while the slope of the mounds remains constant in 
time. The exponent $\beta$ that described this power-law coarsening behavior,
$W(t) \propto R(t) \propto t^\beta$, is found to be close to 0.5. 
The coarsening of mounds in the cusped regime  (i.e. for 
$c_1 > c^{sp}_1(\lambda)$ and any initial configuration, and
$c_1 < c^{sp}_1(\lambda)$ and sufficiently smooth initial configurations)
is qualitatively different. In this growth mode, both the interface width and 
the average slope $s(t)$ of the cusp-like mounds (as well as the maximum 
slope $s_m(t)$) increases with time as a power law in the coarsening 
regime: $W(t) \propto t^\beta$, and $s(t) \propto t^\theta$. This implies 
that the average mound size also increases with time as a power 
law: $R(t) \propto t^n$ with $n=\beta - \theta$. This behavior is 
illustrated in Fig.~\ref{fig7chap4}. The values of the exponents are 
found to be $\beta = 0.54 \pm 0.02$ and $\theta = 0.2 \pm 0.02$, implying 
that the coarsening exponent is $n=0.34 \pm 0.04$. 

In the presence of noise, the steady state behavior is independent
of the initial condition. When the control parameter $c_1$ is
sufficiently large, the nonlinear instability is completely
suppressed and the route to mounding is similar to the linear
instability dominated behavior mentioned above, the configurations
being slightly roughened versions of their noiseless counterpart.
This is illustrated in Fig.~\ref{fig4chap4} where the interface
profiles in a typical run starting from a flat state for
$L=400$, $\lambda=4.0$ and $c_1=0.05$ are shown at times
$t=100$ (early time regime), $t=4000$ (coarsening regime) and
$t=256000$ (steady state). The inset shows the interface width
$W$ as a function of $t$,
averaged over $15$ runs for $L=1000$ samples. The averaged data
show $W \propto t^{\beta}$, $\beta = 0.5 \pm
0.01$ in the coarsening regime.
The average slope of the mounded
interface grows as $s(t) \propto t^\theta$, with $\theta = 0.18 \pm 0.02$
and hence the coarsening exponent is $n = 0.32 \pm 0.03$. 

As the value of $c_1$ is decreased below a critical value holding
$\lambda$ fixed, the nonlinear instability dominates over the linear
one. At early times in runs starting from a flat state,
the interface is self affine and the interface
width shows power law scaling with an exponent close to 0.37. As time
progresses isolated pillars with height $h_0>h_{min}$, where
$h_{min}$ is the minimum height of such a pillar above which the nonlinear
instability is operative, make their appearance through random
fluctuations. The time evolution of the interface beyond the point of
occurrence of the instability is similar to that in the noiseless
situation. In
Fig.~\ref{fig5chap4} we present snapshots of an $L=400$ sample for
$\lambda=4.0$ and $c_1=0.01$ at different stages of growth:
$t=200$ (before the onset of the instability), $t=4000$
(coarsening regime), and $t=2560000$ (steady state). The inset shows
a plot of the interface width as a function of time, obtained by averaging over
$15$ runs for $L=1000$ samples. The averaged data 
show a power-law {\it growth regime} with
exponent $0.37 \pm 0.01$ before the onset of the instability and a
second power-law {\it coarsening regime} with
$W \propto t^{\beta}$, $\beta = 0.49 \pm 0.02$.

A dynamical phase transition at $c_1 =c^{cr}_1(\lambda)
< c^{sp}_1(\lambda)$ separates these two kinds of growth modes. To calculate
$c^{cr}_1(\lambda)$, we start a system at the faceted fixed point and follow its
evolution according to Eq.(\ref{pt_chap4}) for a long time (typically $t=10^4$)
to check whether it reaches a cusped steady state. By repeating this
procedure many times (typically $100$ runs), the probability,
$P_1(\lambda,c_{1})$, of a transition to a cusped state is obtained. For
fixed $\lambda$, $P_1$ increases rapidly from 0 to 1 as $c_1$ is increased above
a critical value. Typical results for $P_1$ as a 
function of $c_1$ for $\lambda=5.0$
are shown in the right inset of Fig.~\ref{fig6chap4}. 
The value of $c_1$ at which
$P_1=0.5$ provides an estimate of $c^{cr}_1$. 
Another estimate is obtained from a
similar calculation of $P_2(\lambda,c_1)$, the probability that a flat initial
state evolves to a faceted steady state. 
As expected, $P_2$ increases sharply from
0 to 1 as $c_1$ is decreased 
(see right inset of Fig.~\ref{fig6chap4}), and the value
of $c_1$ at which this probability is 0.5 is slightly 
lower than the value at which
$P_1=0.5$. This difference reflects finite-time hysteresis effects. 
The value of
$c^{cr}_1$ is taken to be the average of these two estimates, 
and the difference between
the two estimates provides a measure of the uncertainty in the determination of
$c^{cr}_1$. The phase boundary obtained this way is 
shown in Fig.~\ref{fig6chap4}, along
with the results for $c^{sp}_1$.

The scaling behavior in the coarsening regime in the presence of noise is
the same as that found in the noiseless case. The qualitative behavior in
the faceted phase is similar to that found in our earlier work
\cite{Chakrabarti:03,Chakrabarti:04} on this model without the ES term. The ES term, however,
has an important effect: it changes the coarsening exponent from 0.33 to
0.5. A model very similar to the one considered here was studied by
Torcini and Politi~\cite{Torcini:02} 
for parameter values deep in the cusped regime (small $\lambda$,
large $c_1$). The mounded morphology we find in this regime is similar to that
found in their study. Our results for the exponents $\theta$ and $n$ are
slightly different from the values ($\theta = n =1/4$) reported by them. 
This is probably due to crossover effects --
we have found values of $n$ closer to 1/4 if
smaller values of $\lambda$ and/or larger values of $c_1$ are
used. 

In summary we have shown that a nonlinear instability in our spatially discrete
growth model in which the linear ES instability is also present 
may, for appropriate parameter values, lead to 
mound formation with slope selection and power-law coarsening. This is
qualitatively different from the behavior in the parameter regime where the
ES instability dominates: the system exhibits mound formation in this regime
also, but there is no 
slope selection. The coarsening exponent has different values in these
two regimes which are separated by a line of first-order dynamical phase
transitions. The ES part of the surface current in our model does not vanish
for any non-zero value of the slope. Therefore, the slope selection we find
in the regime where the nonlinear instability is dominant is 
qualitatively different from that in ES-type models and is a true example of 
nonlinear pattern formation.  
The noiseless version of our model exhibits an interesting dependence on
initial conditions: the long-time behavior depends on whether the
inhomogeneities in the initial configuration are sufficient to seed the 
nonlinear instability. Both kinds of mounding behavior found in this
study have been observed in experiments, and our model may be relevant 
in the development of an understanding of these experimental observations.
However, more work is required to establish a connection between our results
and those of experiments.


\end{document}